\documentclass{article}

\usepackage{PRIMEarxiv}

\usepackage[utf8]{inputenc}
\usepackage[T1]{fontenc}
\usepackage[numbers,sort&compress]{natbib}
\usepackage{hyperref}
\usepackage{url}
\usepackage{xurl}
\usepackage{booktabs}
\usepackage{amsfonts}
\usepackage{amsmath,amssymb}
\usepackage{nicefrac}
\usepackage{microtype}
\usepackage{fancyhdr}
\usepackage{graphicx}
\usepackage{array}
\usepackage{tabularx}
\usepackage{enumitem}
\usepackage{xcolor}
\usepackage{float}
\usepackage{tikz}
\usetikzlibrary{positioning,arrows.meta,fit,backgrounds,shapes.multipart}

\usepackage{titlesec}
\titlespacing*{\section}{0pt}{1.2ex}{0.8ex}

\title{Organizational Context for the \\ Enterprise Security Brain}

\author{
  \textbf{Eden Yavin},
  \textbf{Ariel Zadok},
  \textbf{Gal Engelberg},
  \textbf{Elad Elouz},
  \textbf{Leon Goldberg}\\
  Sola Security, Tel Aviv, Israel \\
  \texttt{Corresponding author: gal.e@sola.security}
}

\hypersetup{
pdftitle={Organizational Context for the Enterprise Security Brain},
pdfauthor={Eden Yavin, Ariel Zadok, Gal Engelberg, Elad Elouz, Leon Goldberg},
pdfkeywords={organizational context, enterprise security brain, security agents, provenance, knowledge conflicts},
}

\begin{document}

\maketitle

\begin{abstract}
Agents that investigate and decide on enterprise security reason over a machine-readable view of the environment: its assets, identities, configurations, relations, and findings. The correct judgment about an asset also depends on what the organization knows about it: who owns it, what business function it serves, which policy governs it, and which risks were accepted. Part of this knowledge can be inferred from machine state, such as an ownership tag; most of it is recorded by people in policies, audit reports, architecture diagrams, tickets, and conversations, outside machine state. We call this distance the \emph{organizational context gap}. Closing it poses four challenges: extracting the right claims from heterogeneous organizational data and, at inference time, detecting claims made stale by live state, checking live state against the governing policy, and binding architecture diagrams, which carry the business logic, to live resources. We introduce \emph{organizational context} as a layer of the enterprise security brain and define it with a taxonomy and six properties its representation must preserve. We then define a context model in which a claim carries one meaning, an attestation records each source that asserts it, strength is derived from live evidence at read time, and claims resolve onto the entities of the asset graph. On a production environment spanning AWS, Okta, Google Workspace, and GitHub, extraction recovers 0.80 of a pooled 1{,}355-claim ground truth at 0.87 precision, with 0.5\% of extracted claims unsupported by their source. An agent reasoning over live state and resolved context raises the conflict in 0.97 of stale-finding tasks, and reaches answer correctness of 0.83 on policy-alignment tasks and 0.86 on architecture tasks.
\end{abstract}


\section{Introduction}
\label{sec:intro}

Autonomous cyber defense, in which agents assess an organization's security state and act on that assessment with limited human intervention, has become a stated objective of enterprise security programs~\citep{osb}. Yet the security state alone does not determine the correct decision. Our prior work built the structured half of the required context. We call it the \emph{enterprise security brain}: a shared substrate, queried by agents and analysts, whose core is an \emph{asset graph} relating resources, identities, configurations, and findings across vendors, resolved before any question arrives~\citep{assetintel}. Explicit relational context of this kind improves agent answer correctness by approximately 34\% relatively on cross-vendor identity tasks~\citep{crossvendor}. On 28 live cloud investigations, the brain reached a coverage of 0.549 against 0.340 for Claude Code and 0.281 for OpenAI Codex, two general-purpose coding agents with read-only access to the same AWS environment~\citep{brain}. That is 1.61$\times$ and 1.95$\times$ their coverage, at 28.7$\times$ and 40.6$\times$ lower cost per unit of coverage.

The correct judgment about an asset depends on a second kind of fact. Consider a storage bucket flagged as publicly readable at High severity. In one organization it serves static website assets, and the finding is noise. In another it holds payment exports, a signed risk acceptance covers a temporary vendor integration, and no public route reaches it; the correct verdict is a mitigated finding that cites the acceptance. The machine state of the two buckets is identical. The facts that separate them (ownership, business function, data criticality, governing obligations, accepted risk) are held by people and recorded in policies, audit reports, tickets, wiki pages, and conversations. We call the distance between what the environment's state records and what the organization knows about it the \emph{organizational context gap}.

Closing this gap poses four challenges. The first is extraction. Organizational knowledge is spread across heterogeneous sources, the same fact recurs in several of them, and sources differ in authority, so the claims that matter must be recovered from all of this data without dropping requirements or inventing them. The other three arise at inference time. Documented knowledge goes stale: an audit finding can stop holding the moment the infrastructure changes, and language models given conflicting evidence resolve it inconsistently and often defer to retrieved text~\citep{conflictsurvey,stubborn,clasheval}, so a contradiction between a claim and live data has to be represented and surfaced. Compliance requires checking reality against intent: whether a live asset satisfies the specific policy clause that governs it, rather than a plausible clause written for a neighboring asset class. Architecture diagrams complete the picture of the organization. They carry the business logic of the system, which services depend on which and where each one runs, and they become usable only once the components they name are bound to concrete live resources.

We argue that organizational context is a layer of the enterprise security brain with the same standing as the asset graph, and that it should be held as evidence rather than as text. This paper makes three contributions. First, we define organizational context for security reasoning, give a taxonomy of its content, origin, and temporality, and identify six properties that any representation of it must preserve. Second, we define a context model in which a \emph{claim} carries one distinct meaning, an \emph{attestation} records one source asserting that meaning at one point in time, and a \emph{context source} identifies the unit ingested. Claim strength is derived at read time, a single rule governs lifecycle, and claims that resolve to a resource become edges on the asset graph. Third, we place the model in a tiered security brain in which each abstraction is computed once from the tier below, and distinguish context made available to agents from context investigated on top of it.

The evaluation asks whether organizational context, once resolved, lets an agent reach the right judgment about live entities. Resolution means that an assertion, wherever it was recorded, is reduced to a claim, attached to its evidence, placed in time, and bound to the entity it concerns. We examine it from two perspectives. \emph{Extraction} asks whether what the organization asserted is what the brain holds: whether extracted claims cover a document's requirements, add nothing it does not state, and stay faithful to the source. \emph{Inference} asks whether an agent that reasons over live state and resolved context reaches the correct judgment for the right reason. The judgment must rest on the assertion that governs the entity, stay within what the evidence supports, and register disagreement between documented knowledge and live state. We instantiate the three inference challenges as three task sets: stale findings, policy alignment, and architecture diagrams. Both perspectives run on a production environment spanning AWS, Okta, Google Workspace, and GitHub.

The paper proceeds from related work to the definition of organizational context, its place in the security brain, and the context model, followed by the experimental setup and evaluation, results, and discussion.

\section{Related Work}
\label{sec:related}

\textbf{Organizational memory and the company brain.} Organizational memory is the information an organization retains from its past and draws on when it decides in the present~\citep{walshungson}. Walsh and Ungson locate it in individuals, culture, transformations, structures, and ecology; Stein and Zwass add information systems as a retention facility~\citep{steinzwass}. Enterprise knowledge graphs operationalize part of this memory and report entity disambiguation, extraction from unstructured sources, and knowledge evolution as recurring challenges~\citep{noykg}. Agentic AI has renewed the question as a shared knowledge layer that people and agents both read from, often called a company brain. Gartner terms it an enterprise context layer and expects agentic AI efforts to fail without one~\citep{gartnerctx}. Kirchdorfer et al.\ propose such a memory for agents that execute business processes, a governed layer of procedural knowledge shared across agents~\citep{orgmembpm}. A production deployment for enterprise coding agents curates task experience into question--answer memories~\citep{orgmemcoding}. AkasicMEM keeps source authorization in force through transitive lineage~\citep{akasicmem}, and reversible forgetting moves obsolete knowledge between active, dormant, and retired states instead of erasing it~\citep{reversibleforgetting}. General agent memory systems manage a hierarchy of context~\citep{memgpt} or a temporal knowledge graph~\citep{zep}. These systems hold procedural knowledge, agent experience, access policy, and relevance over time. The knowledge we target is declarative and referential: assertions about specific entities, which must be bound to a live asset graph and reconciled with machine state before they change a security judgment.

\textbf{Context layers for security agents.} CRAKEN retrieves from a knowledge base of capture-the-flag writeups during task execution~\citep{craken}. An organization-scoped runtime for regulated security operations carries a typed security context through every retrieval, tool call, memory, and finding~\citep{orgscopedsoc}. Security ontologies and knowledge graphs integrate vulnerability, weakness, and threat knowledge~\citep{uco,sepses}, attack graphs model how exposure composes across hosts~\citep{attackgraphs}, and OSCAL provides machine-readable control catalogs~\citep{oscal}. Our prior work develops the security brain directly, with benchmarks for agentic identity posture management~\citep{ispm,crossvendor}, an open benchmark framework for autonomous cyber defense~\citep{osb}, an asset intelligence layer~\citep{assetintel}, and a comparison with two general-purpose coding agents on live cloud investigations~\citep{brain}. These layers hold threat knowledge, operational scope, and machine-generated state. The asset intelligence layer consumes business-criticality and data-criticality signals; this paper specifies the layer those signals come from when people write them down.

\textbf{Evidence, time, and conflict.} Nanopublications package an assertion with its provenance~\citep{nanopub}, Wikidata attaches references and ranks to each statement~\citep{wikidata}, and W3C PROV standardizes provenance records~\citep{prov}. Bitemporal databases separate when a fact holds from when it was recorded~\citep{bitemporal}. Truth discovery and knowledge fusion estimate source reliability under conflict~\citep{truthdiscovery,knowledgevault}, and entity resolution links records of the same object~\citep{entityresolution}. Benchmarks characterize how models handle real-world contradictions~\citep{wikicontradict} and facts that change over time~\citep{freshqa}. Our context model composes these representational ideas and moves conflict from the model into the data, as explicit relations between claims.

\textbf{Grounding in documents and structured data.} Retrieval-augmented generation conditions generation on retrieved passages~\citep{rag}, agentic variants add planning and tool use~\citep{agenticragsurvey,react}, and graph-based retrieval organizes a corpus into entities and communities~\citep{graphrag,hipporag}, with evaluation frameworks and benchmarks for retrieval quality and faithfulness~\citep{ragas,ares,crag}. Text-to-SQL benchmarks evaluate questions translated into queries over realistic database schemas~\citep{spider,bird,spider2}, and hybrid settings reason over structured and unstructured data together~\citep{hybridqa,stark,tag}. These lines take the passage or the row as the unit of evidence. Our unit is the claim, bound to an entity and carried with its corroboration, authority, and validity.

What is missing across these threads is a representation of what an organization knows about its own environment that preserves corroboration, authority, time, and disagreement, and that resolves onto the live security state agents query. This is the organizational context gap.

\section{Organizational Context}
\label{sec:concept}

Let $E$ be the entities of an enterprise environment (resources, identities, repositories, findings) and $s(e)$ the machine-observable state of $e \in E$: its configuration, metadata, and relations as read from the source systems. A security judgment $J(q,e)$ answers a question $q$ about $e$, such as whether a finding is exploitable, whether $e$ complies with a control, or who must remediate it.

\textbf{Definition.} \emph{Organizational context} is the knowledge of what the entities of an environment mean to the organization (who owns them, what they serve, what data they hold, which obligations govern them, and which risks were accepted) that changes the correct value of some security judgment $J$.

The test is operational: knowledge belongs to organizational context if two identically configured entities would warrant different correct judgments $J(q,e)$ depending on it. Organizational context has two sources. Part of it is encoded in $s$ and must be inferred: a tag \texttt{owner=platform-team}, a naming convention, or an entity's position in the graph indicates ownership or function. The rest is recorded by people outside $s$: the sentence ``we took over prod database operations last quarter'' in a team channel, or the policy clause that requires encryption at rest for regulated data. This work extracts organizational context from the second source and evaluates how it meets $s$ at inference time. The results measure three ways separately: conflict, detecting when $s$ has changed and a documented claim no longer holds (contradiction tasks); compliance, checking the live state of an entity against the claim that governs it (alignment tasks); and resolution, binding the components that an architecture diagram names to the entities $s$ describes (architecture tasks).

Organizational context varies along three axes. By \emph{content}, it asserts ownership and responsibility, business function, data criticality, obligations (policies, standards, controls), exceptions and risk acceptance, or architecture and intent. By \emph{origin}, it is inferred from the connected estate (naming, environment markers, graph position), which is the first source. In the second source, it is asserted directly by a member of the organization, written in a standalone document, or recorded in a ticket, wiki page, or conversation. By \emph{temporality}, it is evergreen (a standing policy), scoped to a window (a finding relevant to one audit period), or ephemeral (a remark in passing). A scoped assertion remains true of its window and relevant to questions about it.

Six properties distinguish organizational context from a corpus of passages, and each imposes a requirement on its representation.

\begin{enumerate}[label=\textbf{P\arabic*},leftmargin=*,itemsep=1pt]
\item \textbf{Redundancy.} The same meaning recurs across sources, and repetition is corroboration. Deleting one source must not delete a meaning that others also assert.
\item \textbf{Unequal authority.} A signed policy, an audit report, a ticket comment, and a remark carry different weight, set by the source type and by the specific source.
\item \textbf{Temporal scope.} Assertions hold over windows: an audit finding concerns its period, an exception expires, a policy version is superseded.
\item \textbf{Disagreement.} Sources revise themselves, contradict one another, and are contradicted by live state. A documented finding goes stale when the infrastructure changes.
\item \textbf{Reference.} Assertions concern specific entities or classes of entities, including an asset type named in a diagram, and are useful only once bound to the entities $s$ describes.
\item \textbf{Provenance under consequence.} A false ``compliant'' verdict masks exposure and a false ``non-compliant'' verdict costs analyst time, so every judgment must be traceable to the exact utterance it rests on, including content that originated as an image or file.
\end{enumerate}

A passage ranked by similarity to a query satisfies P6 in part, through its source metadata. The most similar passage may still be a superseded policy version, a stale finding, or a clause that governs a different asset class.

\section{The Enterprise Security Brain}
\label{sec:brain}

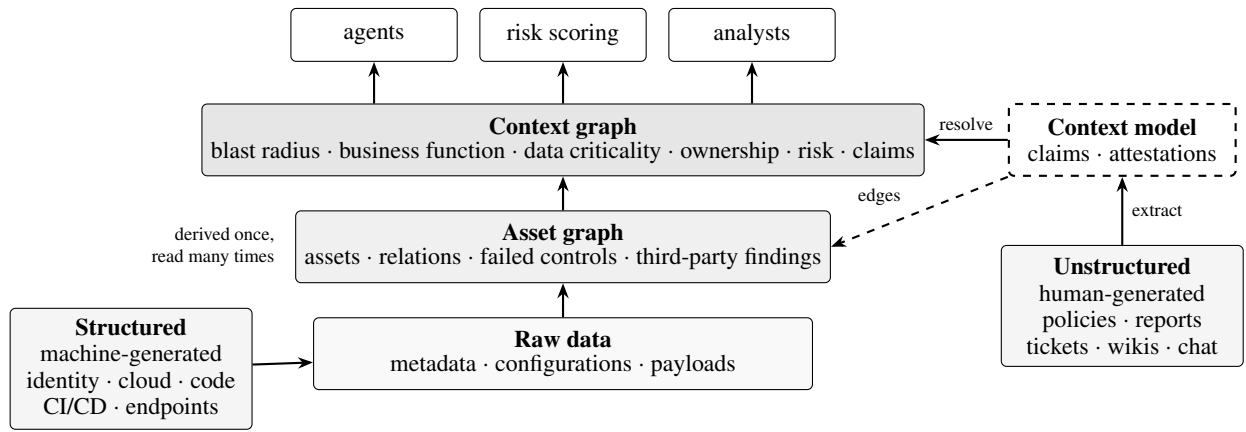
\begin{figure}[H]
\centering
\begin{tikzpicture}[
  tier/.style={draw, rounded corners=2pt, minimum width=6.6cm, minimum height=0.95cm, align=center, font=\small},
  inp/.style={draw, rounded corners=2pt, minimum width=3.2cm, minimum height=1.6cm, align=center, font=\footnotesize, fill=gray!8},
  cons/.style={draw, rounded corners=2pt, minimum width=2.2cm, minimum height=0.7cm, align=center, font=\footnotesize},
  arr/.style={-{Stealth[length=2mm]}, thick},
  ]
  \node[tier, fill=gray!5] (raw) {\textbf{Raw data}\\[-1pt]\footnotesize metadata $\cdot$ configurations $\cdot$ payloads};
  \node[tier, fill=gray!12, above=0.45cm of raw] (ag) {\textbf{Asset graph}\\[-1pt]\footnotesize assets $\cdot$ relations $\cdot$ failed controls $\cdot$ third-party findings};
  \node[tier, fill=gray!20, above=0.45cm of ag] (cg) {\textbf{Context graph}\\[-1pt]\footnotesize blast radius $\cdot$ business function $\cdot$ data criticality $\cdot$ ownership $\cdot$ risk $\cdot$ claims};
  \draw[arr] (raw) -- (ag);
  \draw[arr] (ag) -- (cg);

  \node[inp, left=0.8cm of raw, yshift=-0.2cm] (struct) {\textbf{Structured}\\machine-generated\\identity $\cdot$ cloud $\cdot$ code\\CI/CD $\cdot$ endpoints};
  \node[draw, dashed, thick, rounded corners=2pt, align=center, font=\footnotesize, right=1.1cm of cg, minimum width=3.0cm, minimum height=0.95cm] (ucm) {\textbf{Context model}\\claims $\cdot$ attestations};
  \node[inp, below=0.9cm of ucm] (unstr) {\textbf{Unstructured}\\human-generated\\policies $\cdot$ reports\\tickets $\cdot$ wikis $\cdot$ chat};
  \draw[arr] (struct) -- (raw);
  \draw[arr] (unstr) -- node[right, font=\scriptsize]{extract} (ucm);
  \draw[arr] (ucm.west) -- node[above, font=\scriptsize]{resolve} (cg.east);
  \draw[arr, dashed] (ucm.south west) -- node[above left, font=\scriptsize, pos=0.55]{edges} (ag.east);

  \node[cons, above=0.55cm of cg, xshift=-2.5cm] (c1) {agents};
  \node[cons, above=0.55cm of cg] (c2) {risk scoring};
  \node[cons, above=0.55cm of cg, xshift=2.5cm] (c3) {analysts};
  \foreach \c in {c1,c2,c3} \draw[arr] (cg.north -| \c) -- (\c);
  \node[font=\scriptsize, left=0.15cm of ag, align=right] {derived once,\\read many times};
\end{tikzpicture}
\caption{The enterprise security brain and its organizational context tier.}
\label{fig:brain}
\end{figure}

The enterprise security brain is the substrate that agents, scoring pipelines, and analysts read from when they make security decisions. It is organized as tiers, each computed from the one below (Figure~\ref{fig:brain}). The \emph{raw data} tier holds metadata, configurations, and payloads from source systems. The \emph{asset graph} holds assets, relations, failed controls, and mappings to third-party findings. The \emph{context graph} holds abstractions derived over the asset graph: attack vectors, blast radius, business function, data criticality, and risk scores~\citep{assetintel}, together with ownership and organizational claims. Every tier is resolved before a question arrives, and every tier is accessible at inference time: an agent can read raw records, traverse the asset graph, and query the context graph within one investigation, and security logic runs over the resolved relations when the question arrives~\citep{brain}. Each tier is derived once and read many times, which moves cost into the pipeline and gives every agent and analyst the same answer to the same question. Repeated derivation over raw data yields unstable prioritization~\citep{assetintel}, and live enumeration with a bounded number of turns yields sampled answers to population questions~\citep{brain}.

Organizational context is one more tier of this kind. Its input is largely unstructured knowledge written by people, and its derived form is structured: claims with their evidence, bound to entities of the asset graph wherever a binding exists. The claim ``production databases are managed by the platform team'' becomes an edge on each resource it concerns, next to the technical relations. ``Which crown jewels are internet-exposed'' is then one graph query rather than a document search followed by a manual join against live state. Claims that bind to no entity remain searchable as claims.

The brain exposes organizational context at two levels. At the first level, context is \emph{available}: all organizational context relevant to security is connected, normalized, and exposed through one query surface, as claims with their evidence, to any agent. At the second level, context is \emph{investigated}: use-case-specific reasoning runs over the first level and returns its conclusions as new claims. Severity refinement is one such investigation. An agentic loop reads resource tables, searches organizational context, and walks the graph around a finding, then returns a verdict (mitigated, reduced, amplified, or unchanged) with a confidence, the trace, and the context it cited. Business-function and data-criticality classification is another. Three constraints bound every investigation: it may cite only context it was shown, an investigation that cannot conclude leaves its subject unchanged, and its full path is recorded. Derived claims carry the same provenance and lifecycle as ingested ones and are re-derived when their evidence changes.

\section{The Context Model}
\label{sec:ucm}

The context model rests on one principle: one meaning, one claim, with every occurrence of that meaning kept as evidence.

\begin{figure}[H]
\centering
\begin{tikzpicture}[
  src/.style={draw, rounded corners=2pt, minimum width=3.0cm, minimum height=0.75cm, align=center, font=\footnotesize, fill=gray!8},
  clm/.style={draw, thick, rounded corners=3pt, minimum width=4.4cm, minimum height=1.0cm, align=center, font=\footnotesize},
  res/.style={draw, rounded corners=2pt, minimum width=2.8cm, minimum height=0.8cm, align=center, font=\footnotesize, fill=gray!18},
  att/.style={font=\scriptsize, inner sep=2pt, align=center},
  arr/.style={-{Stealth[length=2mm]}, thick},
  ]
  \node[src] (s1) at (0,2.1) {Security Policy v3\\[-1pt]\scriptsize document};
  \node[src] (s2) at (0,0.6) {\#security-infra\\[-1pt]\scriptsize channel};
  \node[src] (s3) at (0,-0.9) {SEC-1234\\[-1pt]\scriptsize ticket};
  \node[clm] (c) at (7.0,0.6) {\textbf{Claim}\\``Production databases are\\managed by the platform team''};
  \draw[arr] (s1.east) -- node[att, above, sloped, pos=0.45]{\S4.2, p.\,12} (c.west);
  \draw[arr] (s2.east) -- node[att, above, pos=0.45]{permalink, author} (c.west);
  \draw[arr] (s3.east) -- node[att, below, sloped, pos=0.45]{comment \#3} (c.west);
  \node[font=\scriptsize\itshape] at (3.1,2.35) {attestations};
  \node[res] (r) at (12.1,0.6) {\textbf{prod-db-3}\\[-1pt]\scriptsize asset graph};
  \draw[arr] (c.east) -- node[att, above=2pt]{resolves to} (r.west);
  \node[font=\scriptsize, align=center, below=0.35cm of c] {strength derived at read time from\\live attestations: corroboration, authority,\\recency, validity};
\end{tikzpicture}
\caption{Claims, attestations, and context sources. Three sources assert the same meaning; each occurrence is an attestation that carries its own locator, author, time, and authority. The claim resolves to a resource and becomes a queryable edge on the asset graph.}
\label{fig:ucm}
\end{figure}
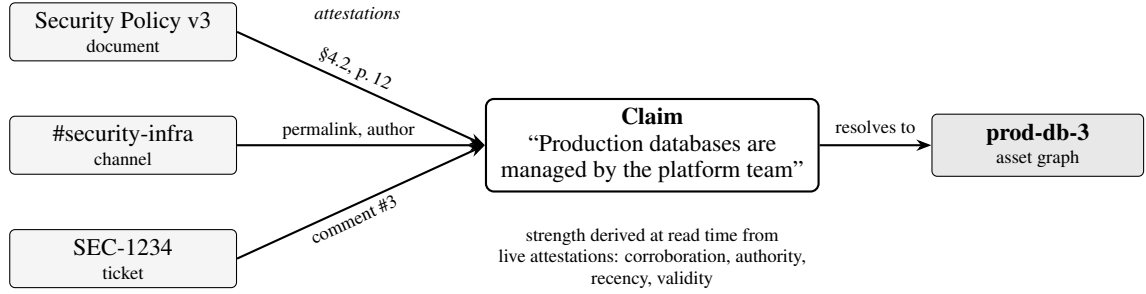

\textbf{Entities.} The model has three entities (Figure~\ref{fig:ucm}). A \emph{claim} is one distinct meaning, independent of any single source, and is the unit of retrieval. An \emph{attestation} records one source asserting a claim at one time, with the verbatim text, author, observation time, validity window, authority, and a \emph{locator}: a chain from the parent source to the exact position of the text, such as a page and section, a ticket comment, or a message permalink. When the content was not native text, the chain ends at the original media object, and a \emph{medium} field records whether it was a file or an image. A \emph{context source} identifies the unit on which ingestion is configured: a ticket, a channel, a page, or a document. Anything true of the meaning lives on the claim; anything true of a particular utterance lives on the attestation. Normalization therefore loses nothing, and every claim can be traced to its quotes, authors, and locations.

\textbf{Strength.} Strength is derived at read time and never stored. Let $A_t(c)$ be the live attestations of claim $c$ at evaluation time $t$:
\begin{equation}
\sigma(c,t) = f\big(n(c,t),\; a(c,t),\; \rho(c,t),\; v(c,t)\big),
\qquad a(c,t) = \max_{\alpha \in A_t(c)} b(\tau_\alpha)\, j(\alpha),
\label{eq:strength}
\end{equation}
where $n$ is the number of distinct sources in $A_t(c)$ (corroboration); $a$ is authority, the product of a configurable base weight $b$ for source type $\tau_\alpha$ and a model judgment $j(\alpha)$ of the specific source (official or informal, senior or unidentified speaker, firsthand or reported); $\rho$ is the newest observation time (recency); and $v$ indicates whether any attestation's validity window covers $t$. We specify the inputs to $f$; its functional form is outside the scope of this paper. The maximum for authority is intentional: one authoritative source suffices, and weak sources contribute through corroboration instead of diluting authority. Validity is a soft signal. A claim attested by a 2024 audit is down-weighted today and still retrievable, so an agent asked about that audit finds it. Since attestations arrive, expire, and disappear with their sources, a stored strength would drift from its evidence.

\textbf{Lifecycle.} A claim lives as long as some source still asserts it. Knowledge therefore ages with its evidence: when every attestation of a claim has passed its validity window, the claim becomes dormant instead of being deleted, remains inspectable, and returns when a source asserts it again. Removing one source never removes a meaning that other sources also hold; only a deliberate retraction removes a fact.

\textbf{Ingestion and resolution.} Every origin, whether a document, a ticket, a conversation, or a direct statement, passes through one path. Each new or revised piece of content is reduced to claims, and each claim is checked against what the brain already knows. A known meaning gains an attestation, which adds to its corroboration; a new meaning becomes a new claim. A claim that revises an earlier one from the same source supersedes it, and the earlier claim is kept as history; a claim that contradicts another source is kept beside it. Each claim is then resolved to the entity, or class of entities, it concerns, and becomes an edge on the asset graph.

\textbf{Retrieval.} Retrieval returns claims, not passages. Each result carries the claim, its derived strength, its attestations with source, locator, authority, and observation time, and its live conflicts:

{\small
\begin{verbatim}
retrieve("who manages prod databases?") ->
{ claim:        "prod DBs are managed by the platform team",
  strength:     high,
  attestations: [ {source: "Security Policy v3", locator: "4.2"},
                  {source: "#security-infra", locator: permalink} ],
  conflicts:    [] }
\end{verbatim}}

An agent that receives a claim with a live conflict, or with attestations outside their validity window, receives the disagreement and the staleness as data.

\section{Experimental Setup and Evaluation}
\label{sec:setup}

The evaluation runs on a live production environment. Amazon Web Services (AWS) hosts compute, storage, Kubernetes workloads, and Identity and Access Management (IAM) principals; Okta is the identity provider; Google Workspace governs collaboration identities and sharing; GitHub holds repositories and branch protection. Tasks often require resolving entities across platforms, such as an Okta identity and the IAM role it assumes.

The organizational corpus has two parts. The first is the organization's governance documents: data classification, data protection, system security, business continuity, and data-at-rest and in-transit policies, SOC~2 control narratives, and audit and penetration-test reports. The second is the real architecture document of the environment under study. It describes the deployed services in figures, including a service dependency graph and a processing pipeline, and in tables that specify each component's compute, sizing, and data stores.

The agent under evaluation is a tool-using investigation agent~\citep{ispm,crossvendor} with two tools: SQL over the live asset graph, and search over the context store, which the extraction pipeline evaluated below populates. Given a task, the agent interleaves both tools and returns a verdict with a justification that cites the live fact and the governing document and section, together with the SQL it issued and the claims it retrieved. The agent model is Claude Sonnet 4.6~\citep{sonnet46}.

Inference is evaluated on three task sets, with examples in Appendix~\ref{app:tasks}. A contradiction task asks whether an IAM role still enables the privilege escalation that a findings report attributes to it. An alignment task asks whether a production security group complies with the network filtering clause of the information security policy. An architecture task asks whether the live queue on an edge of the service dependency graph is configured with the timing that the figure annotates on that edge. Each set starts from a seed set written by a security analyst, with a reference answer that states the outcome and its grounds. A question-generation agent expands the seeds into structured rows, each tied to a target asset, a governing document and section, and an expected verdict, in the spirit of self-instruct~\citep{selfinstruct}. Every row carries a recorded human review outcome (question added, verdict kept, changed, or reverted, asset renamed or repointed, finding retyped) and a note on the check performed, so ground truth is human-verified for every row.

Table~\ref{tab:tasks} lists the metrics reported for each task set. Which source the agent ultimately trusts when report and live state disagree is a deployment preference and is not scored.

Extraction is evaluated against 1{,}355 atomic claims drawn from six documents (internal policies, SOC~2 control narratives, and penetration-test reports). Several frontier models independently extract claims from each document with step-by-step justification, and an arbitrator model resolves their disagreements into one set. The ensemble and the arbitrator are drawn from Claude Sonnet 5~\citep{sonnet5}, Gemini 3.1 Flash-Lite~\citep{geminiflashlite}, and GPT-5.5~\citep{gpt55}. The ground truth is therefore a pool of model outputs, in the sense of pooled test collections in information retrieval~\citep{zobel1998}, rather than an exhaustive human annotation.

Judged metrics use GPT-5.5~\citep{gpt55} and Claude Opus 4.8~\citep{opus48} as judges~\citep{judge}. Extraction is scored on three metrics. \emph{Recall}: for each ground-truth claim, the judge decides whether the extracted set covers it. \emph{Precision}: for each extracted claim, the judge decides whether it matches a ground-truth claim. Both are measured against the pool, so recall is pooled recall: it counts the requirements that at least one ensemble model found, and requirements that no model found are invisible to it~\citep{zobel1998}. \emph{Faithfulness}: each extracted claim is graded against the source text as fully, partially, or not supported, adapting atomic-fact evaluation~\citep{factscore} to a three-level scale.

Retrieval is scored on both sources. \emph{Document retrieval recall} is the fraction of ground-truth document sections the agent retrieved, computed deterministically from source and section metadata. \emph{SQL equivalence} is whether the agent's query is semantically equivalent to a reference query, as judged.

Inference is scored per task set. On contradiction tasks, \emph{conflict detection} scores 1 when the agent raises the conflict between the report and live state. Since every contradiction task contains a conflict, a detection counts as grounded only if the retrieval metrics show that the contradicting evidence was retrieved, which guards against an agent that flags every task. On alignment and architecture tasks, \emph{answer correctness} scores 1.0 when the verdict, cited document, and reasoning are correct, 0.5 when the verdict and live data are correct but the cited document is wrong, and 0.0 otherwise; the middle level isolates a correct verdict grounded in the wrong clause. On alignment tasks, \emph{faithfulness} scores 0.0 when the answer introduces a requirement absent from the retrieved context.

\begin{table}[H]
\centering
\caption{Task sets and their metrics.}
\label{tab:tasks}
\small
\begin{tabularx}{\textwidth}{@{}lr>{\raggedright\arraybackslash}X>{\raggedright\arraybackslash}X@{}}
\toprule
Set & $n$ & Task and ground truth & Metrics \\
\midrule
Contradiction & 30 & Does a named asset still carry a finding reported in a findings document? The live state no longer matches the report in every task. & Conflict detection, document retrieval recall, SQL equivalence \\
\addlinespace
Alignment & 32 & Is a named asset, identity, or artifact aligned with a specific clause of a named policy or SOC~2 control? 16 compliant, 16 non-compliant. & Answer correctness, faithfulness, document retrieval recall, SQL equivalence \\
\addlinespace
Architecture & 16 & Does a live resource conform to what the architecture document specifies for it in a figure or table? The live state matches the document in every task. & Answer correctness, document retrieval recall, SQL equivalence \\
\bottomrule
\end{tabularx}
\end{table}

\section{Results}
\label{sec:results}

Table~\ref{tab:extraction} reports extraction and Table~\ref{tab:inference} reports inference by task set; n/a marks a metric that does not apply to a task set. All scores are in $[0,1]$, and higher is better except for the partially and not supported shares.

\begin{table}[H]
\centering
\caption{Extraction against 1{,}355 ground-truth claims.}
\label{tab:extraction}
\small
\begin{tabular}{@{}lr@{}}
\toprule
Metric & Score \\
\midrule
Recall & 0.80 \\
Precision & 0.87 \\
Faithfulness, fully supported & 0.92 \\
Faithfulness, partially supported & 0.06 \\
Faithfulness, not supported & 0.005 \\
\bottomrule
\end{tabular}
\end{table}

Extraction recovers 0.80 of the pooled ground-truth claims, so one pooled requirement in five is absent from the context store. Precision is 0.87. Of the extracted claims, 0.92 are fully supported by their source, 0.06 partially supported, and 0.005 not supported.

\begin{table}[H]
\centering
\caption{Inference by task set.}
\label{tab:inference}
\small
\begin{tabular}{@{}lccc@{}}
\toprule
Metric & Contradiction ($n=30$) & Alignment ($n=32$) & Architecture ($n=16$) \\
\midrule
Conflict detection & 0.97 & n/a & n/a \\
Answer correctness & n/a & 0.83 & 0.86 \\
Faithfulness & n/a & 0.85 & n/a \\
Document retrieval recall & 0.98 & 0.82 & 0.88 \\
SQL equivalence & 0.91 & 0.78 & 0.75 \\
\bottomrule
\end{tabular}
\end{table}

The agent raises the conflict between a stale finding and live state in 0.97 of contradiction tasks. Retrieval on both sides is the highest of the three sets, with document retrieval recall of 0.98 and SQL equivalence of 0.91, so the detections rest on retrieved contradicting evidence.

Answer correctness on alignment tasks is 0.83 and faithfulness is 0.85, so a minority of answers introduce a requirement the retrieved context does not contain. Document retrieval recall is 0.82, the lowest of the three sets, which is consistent with the governing clause in a policy being harder to locate than a finding in a report.

On architecture tasks, answer correctness is 0.86 and document retrieval recall is 0.88. SQL equivalence is 0.75, the lowest retrieval score across the three sets.

SQL equivalence is below document retrieval recall in every set: 0.91 against 0.98 on contradiction, 0.78 against 0.82 on alignment, and 0.75 against 0.88 on architecture tasks. The gap is widest where the task requires binding a diagram component to a concrete resource.

\section{Discussion}
\label{sec:discussion}

The results answer the four challenges posed in the introduction. First, extraction: the context store holds 0.80 of the pooled requirements, and only 0.5\% of its claims lack support in the source, so its errors are omissions rather than inventions, and every inference result is reached with at least one requirement in five absent. Second, staleness: the agent reports that a documented finding no longer holds in 0.97 of contradiction tasks and retrieves both sides of the disagreement, the report section (0.98) and the live evidence (0.91), because the model keeps a documented claim and the live state side by side. Third, compliance: answer correctness is 0.83, retrieval recall 0.82, and faithfulness 0.85, consistent with errors in locating the clause that governs an asset among similar clauses and in adding requirements the clause does not state. Fourth, architecture: answer correctness on architecture tasks is 0.86, but the live query matches the reference in only 0.75 of them, the lowest score in the evaluation, so binding a drawn component to the resources it denotes is the weakest step. Across all three task sets, the agent finds the governing document more reliably than it writes the query that checks the live entity against it: once organizational context is resolved, the remaining difficulty lies on the live side. We describe this pattern; we do not isolate its cause.

The evaluation is limited in three ways. Its scope is one production environment, 78 inference tasks, and a corpus of documents without tickets or conversations. Its ground truth and grading rely on models: the extraction pool comes from a model ensemble, judged metrics come from language-model judges without expert calibration, and every contradiction task contains a conflict while every architecture task matches its document, so neither set measures the opposite outcome. Its design measures one system with the context model in place, without comparison against alternative representations of organizational context. Future work will extend the evaluation to further environments and to conversational sources, calibrate the judges against expert grading, add control tasks with the opposite outcome, and compare the context model against alternative representations.

\section{Conclusion}
\label{sec:conclusion}

Security judgments depend on knowledge that the organization holds, much of it recorded by people outside machine state. We introduced organizational context as a layer of the enterprise security brain and defined it with a taxonomy and six properties. The context model holds it as claims with evidence, derived strength, one lifecycle rule, and resolution onto the asset graph. On a production environment, extraction recovers 0.80 of pooled ground-truth claims with 0.5\% unsupported, and an agent reasoning over resolved context reaches answer correctness of 0.83 to 0.86 and detects stale documented findings in 0.97 of tasks. The live query is the weakest stage, with SQL equivalence between 0.75 and 0.91. The two open questions are the gain of the context model over alternative representations, and its behavior on conversational sources.

\section*{Acknowledgments}
We thank the Sola security research team for review of the evaluation methodology.

\bibliographystyle{unsrtnat}
\bibliography{references}

\appendix
\section{Example Tasks}
\label{app:tasks}

The questions below are a representative sample of each task set, as posed to the agent, with resource, service, and environment names replaced by placeholders. Architecture tasks span four kinds: diagram conformance, specification conformance, design questions that can be checked live only in part, and inventory.

\textbf{Contradiction tasks.} Questions about findings that a security report documents and the live data is known to contradict.
\begin{enumerate}[leftmargin=*,itemsep=2pt]
\item Please check the $\langle$IAM role$\rangle$ resource in AWS and find if there is a finding about this role enabling privilege escalation by launching compute and attaching an existing role to it.
\item Please check the $\langle$EC2 instance$\rangle$ resource in AWS and find if there is a finding related to the instance metadata service allowing IMDSv1.
\item Please check the $\langle$RDS instance$\rangle$ resource in AWS and find if there is a finding related to the RDS instance having automated backups disabled.
\end{enumerate}

\textbf{Alignment tasks.} Questions that check a live asset against the policy clause that governs it, with compliant and non-compliant ground truths.
\begin{enumerate}[leftmargin=*,itemsep=2pt]
\item Is the production $\langle$security group$\rangle$ in AWS aligned with section 9.3 (IT, Network and System Security) of the Information Security Policy?
\item Is the $\langle$security group$\rangle$ in AWS aligned with section 9.3's network traffic filtering requirement of the Information Security Policy?
\item Is the $\langle$S3 bucket$\rangle$ in AWS aligned with the Data Protection policy?
\item Is the production $\langle$application load balancer$\rangle$ in AWS aligned with section 9.6's communication protocols requirement of the IT, Network and System Security policy?
\end{enumerate}

\textbf{Architecture tasks}, one per kind. Questions that check live resources against what the architecture document specifies for them; the live state matches the document in every task.
\begin{enumerate}[leftmargin=*,itemsep=2pt]
\item \emph{Diagram conformance.} Per the service architecture document, Figure 1 calls out queue timing settings on the edge feeding $\langle$service$\rangle$; in the $\langle$environment$\rangle$ environment, is the actual queue instance behind that edge configured the way the diagram says?
\item \emph{Specification conformance.} Per the service architecture document, the Components table pins Fargate sizing for $\langle$service$\rangle$; in the $\langle$environment$\rangle$ environment, is the live task definition behind that service actually provisioned that way?
\item \emph{Design with partial live reach.} Does $\langle$service$\rangle$ scale based on CPU/memory utilization or another metric, and is the live service in the $\langle$environment$\rangle$ environment sized and queued that way?
\item \emph{Inventory.} List all the components in the service architecture that are deployed as AWS Lambda functions, and confirm they exist in the $\langle$environment$\rangle$ environment.
\end{enumerate}

\end{document}